\documentclass[twocolumn,aps,prl,nopacs,superscriptaddress]{revtex4}
\usepackage[latin9]{inputenc}
\usepackage{mathtools}
\usepackage{amsmath}
\usepackage{graphicx}

\makeatletter

\DeclareFontEncoding{LGR}{}{}
\DeclareRobustCommand{\greektext}{%
  \fontencoding{LGR}\selectfont\def\encodingdefault{LGR}}
\DeclareRobustCommand{\textgreek}[1]{\leavevmode{\greektext #1}}
\ProvideTextCommand{\~}{LGR}[1]{\char126#1}

\@ifundefined{textcolor}{}
{%
 \definecolor{BLACK}{gray}{0}
 \definecolor{WHITE}{gray}{1}
 \definecolor{RED}{rgb}{1,0,0}
 \definecolor{GREEN}{rgb}{0,1,0}
 \definecolor{BLUE}{rgb}{0,0,1}
 \definecolor{CYAN}{cmyk}{1,0,0,0}
 \definecolor{MAGENTA}{cmyk}{0,1,0,0}
 \definecolor{YELLOW}{cmyk}{0,0,1,0}
}

\usepackage{verbatim}
\usepackage{sidecap}
\usepackage{epstopdf}
\usepackage{braket}
\usepackage{soul}
\usepackage[labelsep=period]{caption}

\makeatother

\begin{document}
\title{Low energy band structure and symmetries of UTe$_{2}$ from angle resolved photoemission spectroscopy}
\author{Lin Miao}
\thanks{These authors contributed equally to this work}
\affiliation{School of Physics, Southeast University, Nanjing 211189, China.}
\author{Shouzheng Liu}
\thanks{These authors contributed equally to this work}
\author{Yishuai Xu}
\author{Erica C. Kotta}
\affiliation{Department of Physics, New York University, New York, New York 10003, USA}
\author{Chang-Jong Kang}
\affiliation{Department of Physics and Astronomy, Rutgers University, Piscataway, NJ, 08854-8019, USA}
\author{Sheng Ran}
\affiliation{NIST Center for Neutron Research, National Institute of Standards and Technology, 100 Bureau Drive, Gaithersburg, MD 20899, USA}
\affiliation{Quantum Materials Center, Department of Physics, University of Maryland, College Park, MD 20742, USA}
\author{Johnpierre Paglione}
\affiliation{Quantum Materials Center, Department of Physics, University of Maryland, College Park, MD 20742, USA}
\author{Gabriel Kotliar}
\affiliation{Department of Physics and Astronomy, Rutgers University, Piscataway, NJ, 08854-8019, USA}
\author{Nicholas P. Butch}
\affiliation{NIST Center for Neutron Research, National Institute of Standards and Technology, 100 Bureau Drive, Gaithersburg, MD 20899, USA}
\affiliation{Quantum Materials Center, Department of Physics, University of Maryland, College Park, MD 20742, USA}
\author{Jonathan D. Denlinger}
\email{jddenlinger@lbl.gov}
\affiliation{Advanced Light Source, Lawrence Berkeley National Laboratory, Berkeley, CA 94720, USA}
\author{L. Andrew Wray}
\email{lawray@nyu.edu}
\affiliation{Department of Physics, New York University, New York, New York 10003, USA}

\begin{abstract}

The compound UTe$_{2}$ has recently been shown to realize spin triplet superconductivity from a non-magnetic normal state. This has sparked intense research activity, including theoretical analyses that suggest the superconducting order parameter to be topologically nontrivial. However, the underlying electronic band structure is a critical factor for these analyses, and remains poorly understood. Here, we present high resolution angle resolved photoemission (ARPES) measurements covering multiple planes in the 3D Brillouin zone of UTe$_{2}$, revealing distinct Fermi-level features from two orthogonal quasi-one dimensional light electron bands and one heavy band. The electronic symmetries are evaluated in comparison with numerical simulations, and the resulting picture is discussed as a platform for unconventional many-body order.

\end{abstract}
\maketitle
\date{\today}

\begin{figure}
\includegraphics[width=8cm]{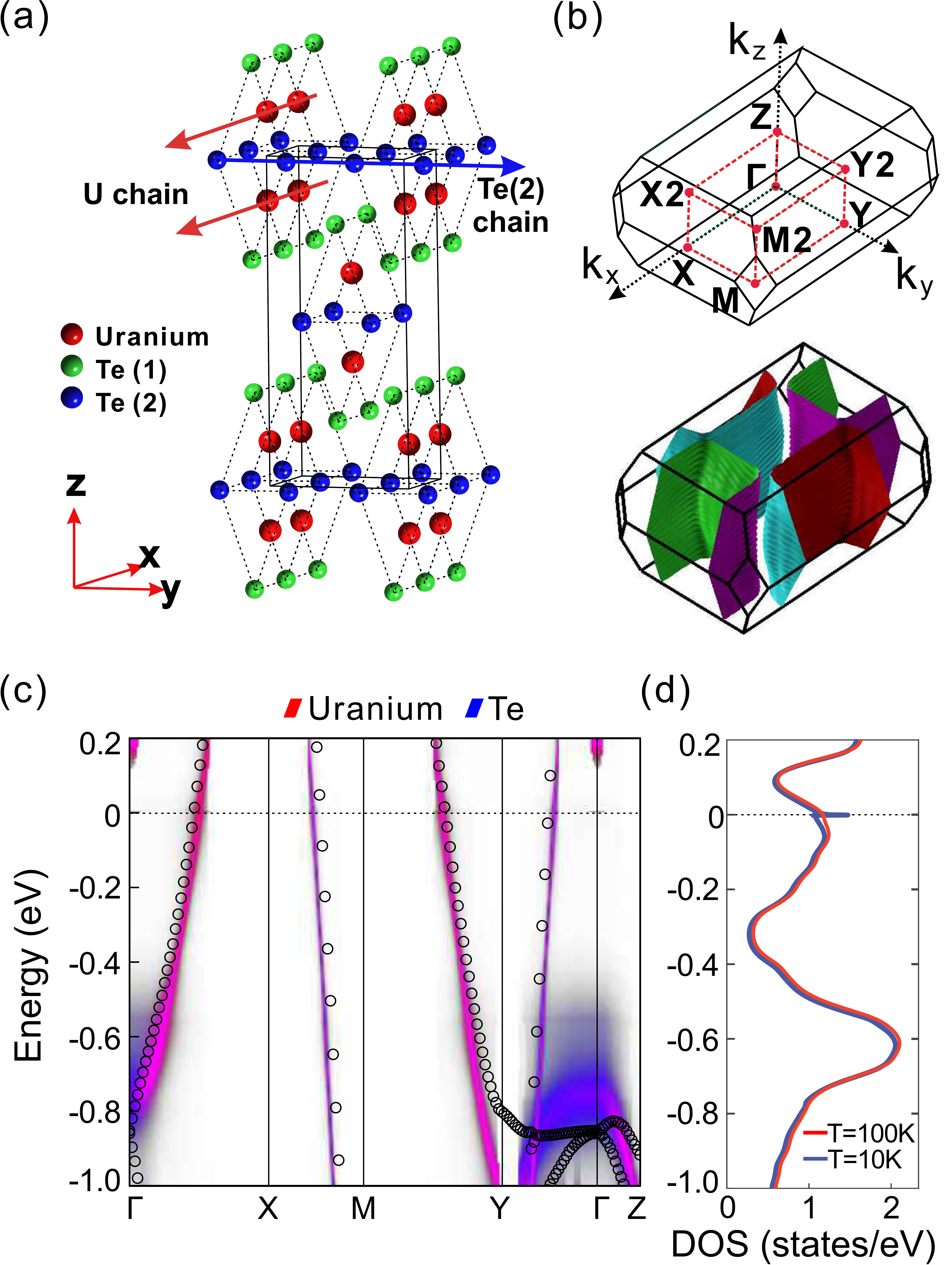} \captionsetup{font={small}}
\captionsetup{justification=raggedright,singlelinecheck=false}
\caption{\textbf{Quasi-1D sublattices of UTe$_2$.} (a) The crystal structure of orthorhombic UTe$_2$. The red arrow labels the uranium chain along a-axis. The blue arrow labels the Te(2) chain along the b-axis. (b) The first Brillouin zone of UTe$_2$ and the DFT-calculated Fermi surface for Th-substituted ThTe$_2$, showing the two light bands. (c) The non-$\it{f}$-orbital DFT band structure (empty circles) is compared with the atomic sites-resolved DMFT+DFT band structure (ribbons) of UTe$_2$ along k-space paths traced with dashed lines in panel (b). Red and blue shading represent the partial density of states from uranium and tellurium orbitals, respectively. The DFT bands are shifted upwards by 200 meV, for better correspondence with the experimental band structure. (d) The DMFT calculated density of states (DOS) at temperatures of T=10K and 100K, below and above the onset of Kondo coherence.}
\end{figure}

Recent years have seen a rapid growth in research on triplet-like superconductivity, driven in part by proposed links to Majorana fermion-based quantum information storage \cite{1}. The superconducting state of the heavy fermion compound uranium diteluride (UTe$_{2}$) has recently been proposed as a promising and potentially unique example of such an order parameter emerging from a non-magnetic normal state \cite{2,3}. However, first principles calculations have predicted a very wide range of underlying band structures \cite{3,4,5,6,7}, due in part to the modeling complexity associated with $\it{f}$-electron strong correlations and Kondo lattice physics. Recent soft X-ray ARPES measurements have provided a first look at the electronic structure, but are limited in energy resolution and by strong incoherent scattering at the Fermi level, and have not established a straightforward agreement with numerical simulations \cite{4}. In this letter, we present high resolution angle resolved photoemission (ARPES) measurements of the electronic band structure of UTe$_{2}$ covering much significant planes in the 3D Brillouin zone. Highly dispersive Fermi level features are shown to correspond closely with first-principles-based simulations combining density functional theory and dynamical mean field theory (DFT+DMFT). These rectangular Fermi pockets originate from two orthogonal one-dimensional (1D) bands, only one of which has strong uranium character. Non-dispersive `heavy' band features associated to the 5$\it{f}$-orbital of uranium are also discovered, with strong implications for many-body ordering instabilities.

Triplet-based superconductivity is strongly implicated in related uranium compounds such as UGe$_{2}$, UReGe and UCoGe \cite{8,9,10}, where the transition to superconductivity occurs within a ferromagnetic normal state. While UTe$_{2}$ does not host long range magnetic order, the attribution of a non-singlet order parameter comes from similar factors. For example, the strongly anisotropic upper critical field (H$_{c2}$) of the UTe$_{2}$ superconducting phase is as high as 35T, which exceeds the Pauli limit for a singlet superconducting pair \cite{11,12}, and the Knight shift is anomalously constant through the superconducting transition \cite{2}. The phase diagram under high magnetic fields depicts a regime in which superconductivity can be field-stabilized \cite{11,13,14,15}. The evidence of thermal transport, heat capacity, and magnetic penetration depth measurements suggest point-like nodal structure that likewise corroborates a triplet superconducting picture \cite{16}. Unconventional Cooper pairing is further suggested by surface probes \cite{17,18} and by measurements showing very strong magnetic fluctuations that coexist with superconductivity \cite{19}, and appear to play a role in enhancing the superconducting critical temperature \cite{20}.

\begin{figure}
\includegraphics[width=7.5cm]{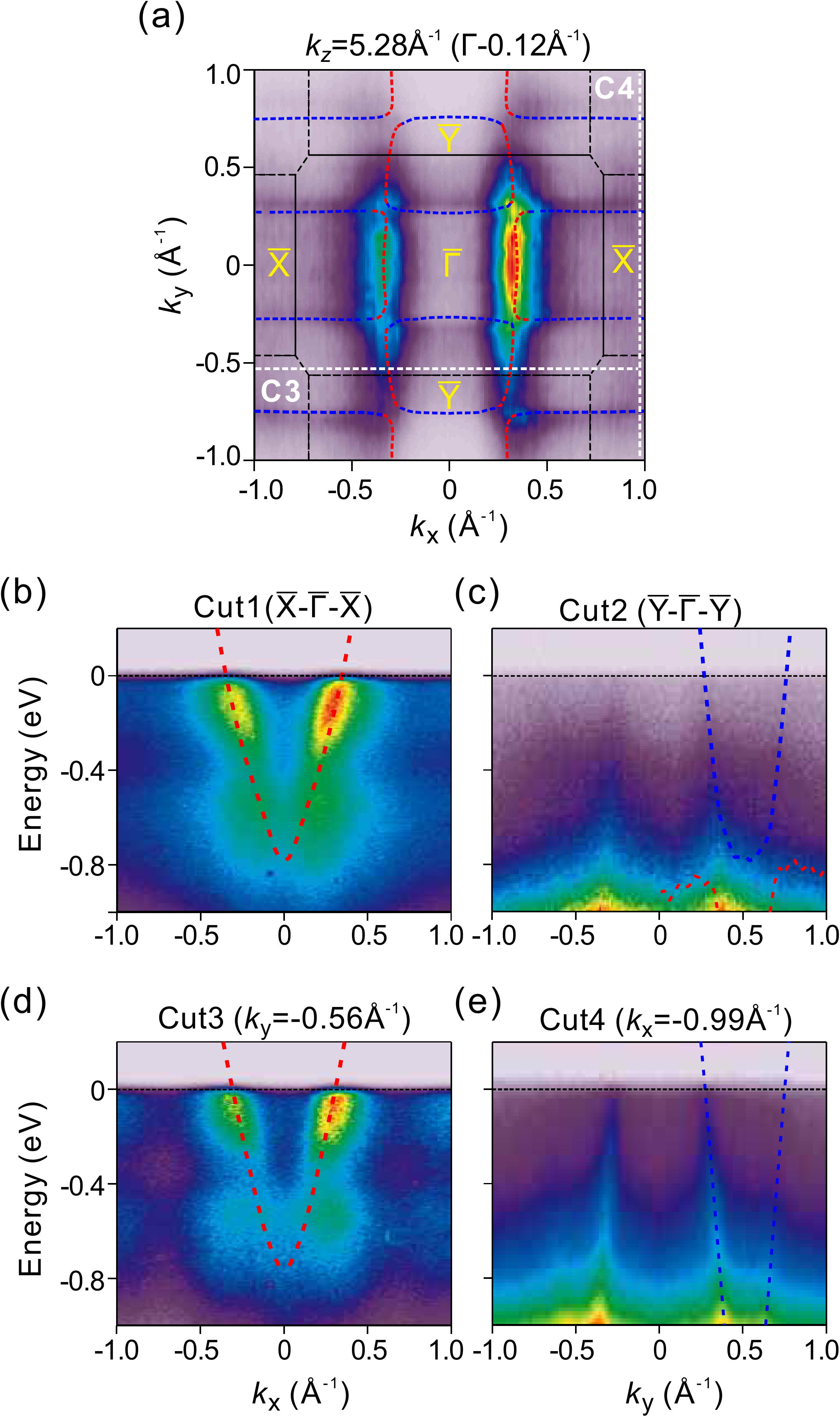} \captionsetup{font={small}}
\captionsetup{justification=raggedright,singlelinecheck=false}
\caption{\textbf{A quasi-1D by quasi-1D light-band Fermi surface.} (a) The ARPES Fermi surface of UTe$_2$ at the (001) crystal face. All panels are overlaid with the DFT band structure, with the quasi-1D light uranium band drawn in red, and blue lines showing the orthogonally dispersing Te(2) band. (b-c) ARPES measurements through 2D Brillouin zone center along the (b) k$_{x}$ and (c) k$_{y}$ directions. (d-e) Additional measurements that are offset from the Brillouin zone center. The trajectories for these cuts are labeled C3 and C4, respectively, on panel (a). Spectra in panels (a,b,d) were measured on the uranium $\it{O}$$_{4,5}$-edge resonance (h$\nu$=98 eV) to enhance visibility for the U band. Panels (c,e) were measured off resonance (h$\nu$=92 eV), and do not include a DFT overlay on the left hand side for visual clarity.\quad{}}
\end{figure}

In this study, ARPES measurements were performed at the MERLIN ARPES endstation beamline 4.0.3 at the Advanced Light Source, and using a helium lamp light source at NYU. The temperature was maintained at T=20K, and the base pressure was similar to 5$\times$10$^{-11}$ Torr. Samples were prepared in a nitrogen glove box and transferred rapidly to ultra-high vacuum (UHV) for in situ cleavage. After the ARPES experiment, the cleaved sample surface was characterized by X-ray Laue diffraction and the microscope, confirming the (001) cleavage surface (see Supplemental Material, SM \cite{21}, Fig. S1). The strongly correlated electronic structure of UTe$_{2}$ is modeled using first principles-based dynamical mean field theory (DFT+DMFT) \cite{22,23,24,25}. The vertex corrected one-crossing approximation [23] was chosen as the impurity solver, in which full atomic interaction matrix was taken into account \cite{25}. The Coulomb interaction U = 6.0 eV, the Hund$'$s coupling J = 0.57 eV, and a nominal double counting scheme were used for the DFT+DMFT calculations. A simplified DFT band calculation was performed for comparison, replacing uranium with thorium within the UTe$_{2}$ crystal structure. This removes 5$\it{f}$ orbitals from the picture near the Fermi level, as thorium strongly favors a vacant quadrivalent thorium 5$\it{f}$$^0$ state. Additional discussion of this scenario is provided in the online supplement \cite{21} (see Fig. S2-S4). The DFT calculations in both methods were performed using the full-potential linearized augmented plane-wave (FLAPW) band method, as implemented in the WIEN2K package \cite{26}.

\begin{figure*}
\includegraphics[width=17cm]{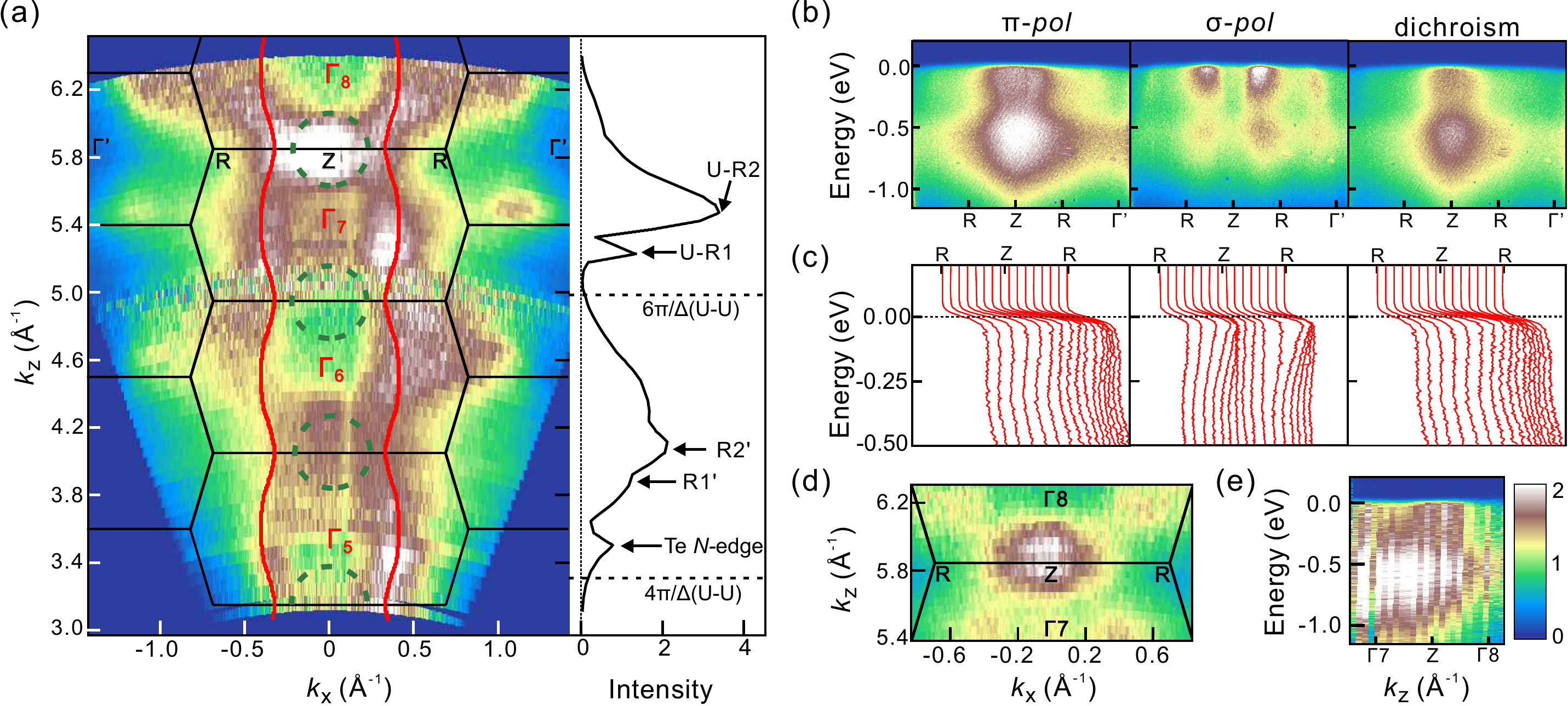} \captionsetup{font={small}}
\captionsetup{justification=raggedright,singlelinecheck=false}
\caption{\textbf{3D dispersion and a heavy Z-point electron pocket.} (a) The k$_{x}$ - k$_{z}$
Fermi surface of UTe$_{2}$, measured by tuning incident energy from h$\nu$=30
to 150 eV (\textgreek{p}-polarization). The approximate contour of an electron pocket at the Z-point is traced in green, and a red curve shows the non-$\it{f}$-orbital DFT-calculated uranium band. The k$_x$-integrated intensity at the Fermi level is plotted at right, and has been divided out from the image at left for visual clarity. Resonance energies annotated on the intensity curve include the R1 and R2 resonances of the uranium $\it{O}$$_{4,5}$-edge, the 40 eV tellurium $\it{N}$-edge, and incoherent background from higher harmonic light intersecting uranium $\it{O}$$_{4,5}$-edge (R1\textquoteright , R2\textquoteright). (b) Polarization-resolved measurements of the Z-point (k$_{z}$$\sim$5.9{\AA}$^{-1}$, h$\nu$=125eV) showing (left) \textgreek{p}-polarization, (middle) \textgreek{sv}-polarization, and (right) the dichroic difference (A(\textgreek{p})-A(\textgreek{sv})). (c) Raw data energy dispersion curves (EDC) for panel (b). (d) A dichroically subtracted Z-point Fermi surface map (A(\textgreek{p})-A(\textgreek{sv})). (e) The Z-point feature seen in E-k$_{z}$ dispersion from the panel (a) data set.}
\end{figure*}

The UTe$_{2}$ lattice is orthorhombic, and belongs to the 71-Immm space group \cite{27}. As shown in Fig. 1(a), the lattice hosts two significant chain-like structures along orthogonal axes. Uranium atoms appear as dimers that are closely separated along the z-axis, and are organized in chains parallel to the x-axis, with a trigonal prism of two chemically inequivalent near-neighbor Te sites, labeled Te(1) and Te(2). The plane of Te(2) atoms also form linear chains along the y-axis with small nearly uniform 3.1{\AA} separation, suggesting large hopping mobility along this axis, particularly when compared to the far-larger next neighbor Te-Te separation of 4.2{\AA} that occurs along the x-axis.

As a first step to understand the electronic structure, we review the simplified DFT simulation in which uranium is replaced with thorium in the crystal to obtain the non-$\it{f}$-electron bands. The resulting Fermi surface consists of four rectangular Fermi pockets, formed by two hybridized bands that have strong 1D character along the k$_x$ and k$_y$ axes (Fig. 1(b)). The simplified DFT calculation is overlaid on the full DFT+DMFT spectral function in Fig. 1(c), revealing an extremely close visual correspondence. Scrutinizing dispersions near the Fermi level, one can see that the band dispersing along k$_{x}$ ($\Gamma$-X) has primarily uranium character (red shading), while the band dispersing along k$_{y}$ ($\Gamma$-Y) is associated with Te orbitals (blue-purple shading), in keeping with expectations from the crystal structure analysis. A more detailed orbital decomposition is presented in the online supplementary information ([21], Fig. S3), showing that U 6$\it{d}$$_{z^2}$ character dominates the electron-like 1D dispersion along k$_{x}$ and a strong bonding-antibonding splitting of the Te(2) 5$\it{p}$$_{z}$ orbitals is the origin of the hole-like 1D Fermi surface sheets. The 5$\it{f}$ bands are removed from the Fermi level by the large Hubbard U value, a phenomenon also noted as possible in earlier numerical simulations \cite{6,7}. However, examining the density of states (DOS) reveals a small peak at the Fermi level representing the emergent band structure from Kondo coherence. Though this band is highly significant in defining the low temperature physics of the material, it is not describable by DFT \cite{28}, and is essentially invisible within the momentum-resolved electron annihilation spectral function probed by ARPES.

Overlaying the DFT result on an experimental k$_{x}$ - k$_{y}$ plane Fermi surface seen by ARPES at the uranium $\it{O}$-edge resonance (h$\nu$=98 eV) reveals a very similar checkered structure (Fig. 2(a)). In this and other comparisons, the non-$\it{f}$-orbital DFT bands are shifted upwards by 200 meV to enhance correspondence with the experimental data. Dispersion measurements performed on resonance show a light band dispersing along k$_{x}$ with a Fermi velocity that is roughly 50$\%$ of the DFT calculation (Fig. 2(b, d)), and a broad non-dispersive feature roughly at 0.7 eV binding energy that has also been noted in an earlier study \cite{4}. The band that disperses along the k$_{y}$ axis has a negative slope, and closely matches the calculation with a remarkably large band velocity of v$_{F}$$\sim$10 eV$\cdot${\AA} (Fig. 2(c,e)), that results from the very large 8 eV bonding-antibonding splitting of the Te(2) linear chain $\it{p}$$_{z}$ orbitals (see SM \cite{21}, Fig. S4). The quasi-1D nature of these bands at the Fermi level is consistently observed at a range of k$_{z}$ coordinates in momentum space (see SM \cite{21}, Fig. S5).

To better understand the three dimensional electronic structure, k$_{z}$-axis momentum dependence of the Fermi surface is mapped in Fig. 3(a) over the 5$^{th}$ to 8$^{th}$ Brillouin zones ($\Gamma$$_{5}$-$\Gamma$$_{8}$). Overlaying the light uranium band from the non-$\it{f}$-orbital DFT calculation (red curve) reveals a clear periodic correspondence, and yields a very standard inner potential value of V$_{0}$=13 eV for ARPES k$_{z}$ calibration. Disregarding core level resonances (labeled on Fig. 3(a, right)), the ARPES intensity shows minima and maxima that are approximately periodic with the U-U dimer separation of $\Delta$(U-U)=3.79 {\AA} (see dashed lines). Intensity is suppressed at k$_{z}$=n$\cdot$2$\pi$/$\Delta$(U-U) (for integer n) as expected for a band that is largely antisymmetric for reflections of the z-axis across the Te(2) plane.

Much of the k$_{z}$ Fermi surface image in Fig. 3(a) is dominated by a strong and inhomogeneous background from incoherent processes, as is common for incident energy maps in the extreme ultraviolet. Sharp band features are most visible in the 5$^{th}$ and 7$^{th}$ Brillouin zones. Focusing on these regions reveals that the Z-point carries enhanced spectral weight (near k$_{z}$$\sim$4.1{\AA}$^{-1}$ and 5.9{\AA}$^{-1}$), suggestive of an electronic state that is not visible in the predicted spectral function. This feature has strong polarization dependence, and is almost invisible in a $\sigma$-polarization ARPES measurement (Fig. 3(b, center)). Taking the dichroic difference in Fig. 3(b, right) reveals that it resembles a shallow non-dispersive (heavy) state. Examining the k$_{z}$-axis suggests that the feature may be roughly isotropic in the k$_{x}$-k$_{z}$ plane (see Fig. 3(d,e)), but the large energy width and highly non-dispersive nature make it difficult to make a fine determination.

Though the light bands traced in Fig. 2-3 have clean attributions, the more non-dispersive features seen at the Z-point Fermi level and at E=-0.7 eV do not correspond with prominent features in the calculations. Though DFT+DMFT predicts a DOS maximum with predominant Te character predicted at a similar energy, it does not overlap with these states in momentum space. To better understand this, we examine the difference between measurements performed on- and off-resonance at the uranium $\it{O}$-edge (Fig. 4(a-b)). The $\it{O}$-edge ARPES spectrum clearly enhances final states associated with the light 6$\it{d}$ uranium band and at $\sim$0.7eV binding energy, however the heavy Z-point band is not visible as the z-axis momentum is far from the Z-point. The E=-0.7 eV feature is specifically enhanced at low in-plane momentum (k$_{x}$ $<$ 0.5{\AA}$^{-1}$) and has a clear sub-structure, with two features identifiable in the momentum-integrated intensity, separated by 0.2 eV (Fig. 4(b), red markers). An $\it{O}$-edge resonant feature with roughly the same k$_{x}$-resolved intensity profile is visible in ARPES on URu$_{2}$Si$_{2}$ \cite{29}, and 5$\it{f}$$^2$ atomic multiplet excitations at $\sim$0.7eV are the most prominent high energy inelastic features in $\it{O}$-edge resonant inelastic X-ray scattering (RIXS) from 5$\it{f}$$^2$ uranium at this photon energy \cite{30}. Examining the DFT+DMFT simulation (Fig. 4(c)), we find that two 5$\it{f}$$^2$ atomic multiplet configurations with $^3$H$_5$ and $^3$F$_2$ symmetry are centered on $\sim$0.7 eV binding energy and separated by 0.2 eV, providing a likely explanation for the feature. If this attribution is correct, it strongly supports the picture presented by the numerics of a $^3$H$_4$-based Kondo lattice.

\begin{figure}
\includegraphics[width=8cm]{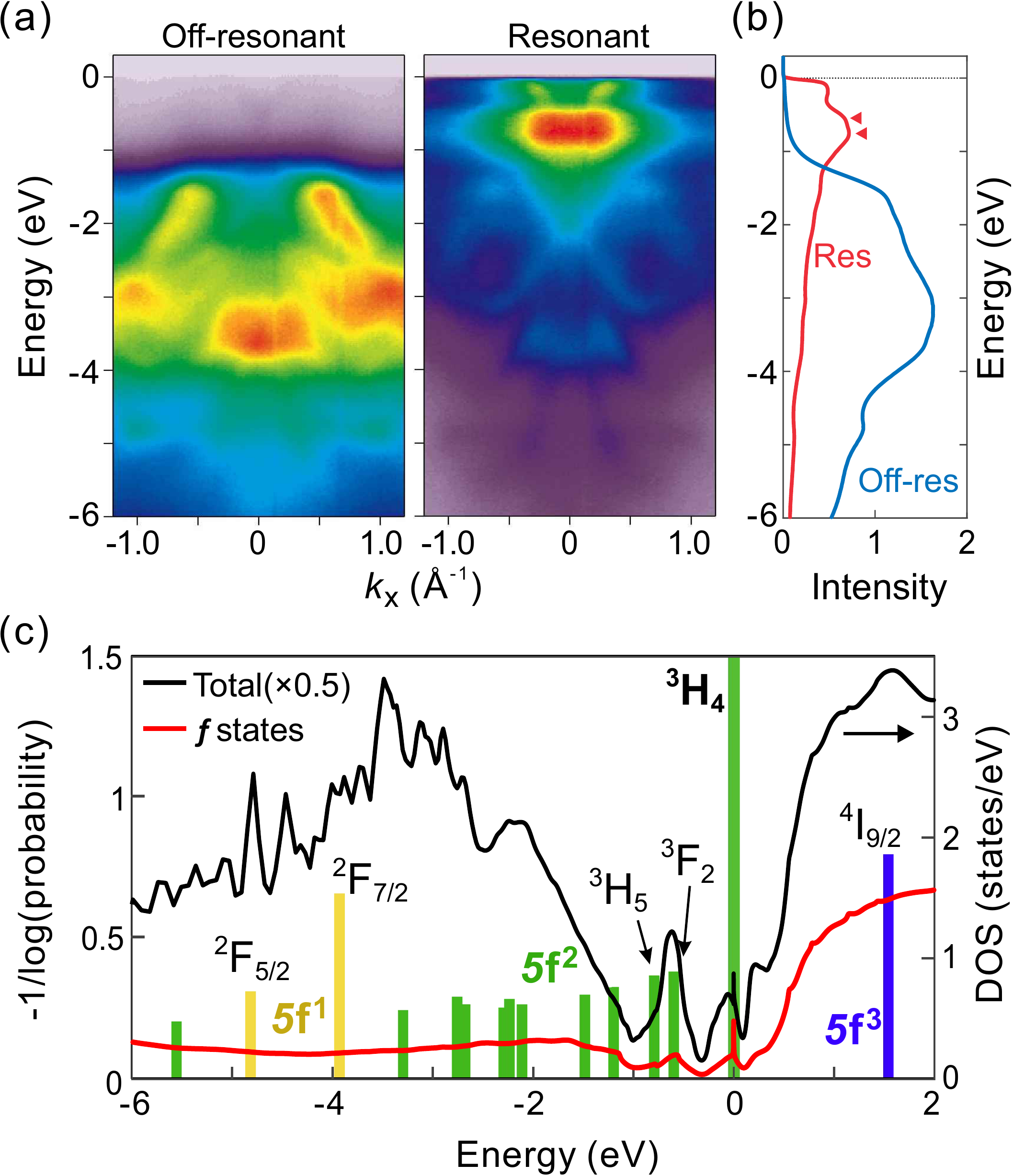} \captionsetup{font={small}}
\captionsetup{justification=raggedright,singlelinecheck=false}
\caption{\textbf{Comparison between resonant ARPES and off-resonant ARPES.}(a) (left) An off-resonance ARPES spectrum with h$\nu$=92 eV is compared with (right) an on-resonance measurement at h$\nu$=98 eV (uranium $\it{O}$$_{4,5}$-edge). (c) Momentum-integrated intensity is shown from data in panel (a). These measurements used a different polarization condition than Fig. 2(b-c) (linear horizontal vs. circular). Red markers identify dual peaks within the $\sim$0.7eV binding energy feature. (c) Density of states (DOS) curves from DFT+DMFT are overlaid with bars representing the occupancy of $\it{f}$-orbital atomic multiplet configurations. The 5$\it{f}$$^2$ Hund$\textquoteright$s rule ground state is the dominant configuration ($^3$H$_4$, 84$\%$ occupancy). The 5$\it{f}$$^1$ and 5$\it{f}$$^3$ configurations would be associated with lower and upper Hubbard bands, respectively, in the absence of other overlapping bands and interactions.}
\end{figure}

The heavy Z-point Fermi-level feature has an even more direct significance for correlated physics \cite{17,27,21}, and may also provide an important channel for spin-triplet Cooper pairing. Simulations suggest the strongest magnetic interactions between uranium atoms to be ferromagnetic coupling within the uranium dimer, with an energy scale that may rise to tens of millielectron volts \cite{7}. This suggests that triplet-favoring ferromagnetic coupling might be strongest for electrons intersecting on the same dimer. However, Pauli exclusion largely forbids this intersection of spin-aligned electrons if only a single itinerant uranium band is present at the Fermi level, as there will be just one associated Wannier orbital shared by the dimer atoms. The existence of a second uranium band at the Fermi level is thus a prerequisite for strong interactions through this channel, and is fulfilled by the observation of the Z-point pocket. Though we cannot provide a definitive attribution for this Z-point feature from theory, the distribution of intensity is consistent with a shallow electron pocket (see band overlay in SM \cite{21}, Fig. S7). The incident energy dependence of this feature is suggestive of odd z-axis reflection symmetry within the uranium dimer, and strong suppression under $\sigma$- (x-axis) polarization near the k$_x$=0 high symmetry plane implies predominantly odd x-axis reflection symmetry.

In summary, we present the electronic band structure study of triplet-like superconductor UTe$_2$ by high-resolution ARPES. Measurements reveal two light quasi-one dimensional bands at the Fermi level, that are attributed to uranium and Te(2) chains through an analysis of resonance and dispersion, as well as a comparison with band calculations. A heavy electronic band is observed surrounding the Z-point with predominantly odd reflection symmetry along the lattice x- and z-axes, representing an important constituent for heavy Fermion physics. A non-dispersive feature at $\sim$0.7 eV binding energy is associated with excitations of a uranium $^3$H$_5$ and $^3$F$_2$ atomic multiplet Kondo lattice, through comparison with DFT+DMFT and experimental data on other systems. The significance of this electronic structure for many-body correlations is discussed, and a favorable channel for triplet-like Cooper pairing is proposed.

\section*{ACKNOWLEDGEMENTS}

This research used resources of the Advanced Light Source, which is a DOE Office of Science User Facility under contract no. DE-AC02-05CH11231. Work at NYU was supported by the MRSEC Program of the National Science Foundation under Award Number DMR-1420073. Synthesis and analysis instrumentation at NYU is supported by NSF under MRI-1531664, and by the Gordon and Betty Moore Foundation's EPiQS Initiative through Grant GBMF4838. Research at the University of Maryland was supported by the Department of Energy, Office of Basic Energy Sciences under Award No. DE-SC0019154, and the Gordon and Betty Moore Foundation's EPiQS Initiative through Grant No. GBMF4419. Identification of commercial equipment does not imply recommendation or endorsement by NIST.

\end{document}